\documentclass[preprint]{emulateapj}

\def\Mstar{M_{\rm star}}
\def\rhostar{\rho_{\rm star}}
\def\Msun{M_{\odot}}

\shorttitle{The Lifetime of Short Gamma-Ray Burst Progenitors}
\shortauthors{Zheng \& Ramirez-Ruiz}

\begin{document}

\title{Deducing the Lifetime of Short Gamma-Ray Burst Progenitors from
       Host Galaxy Demography}

\author{ Zheng Zheng\altaffilmark{1,2} and Enrico
         Ramirez-Ruiz\altaffilmark{1,3} } \altaffiltext{1}{ Institute
         for Advanced Study, Einstein Drive, Princeton, NJ 08540;
         zhengz,enrico@ias.edu} \altaffiltext{2}{Hubble Fellow}
         \altaffiltext{3}{Chandra Fellow}

\begin{abstract}
The frequency of short gamma-ray bursts (GRBs) in galaxies with
distinct star formation histories can be used to constrain the
lifetime of the progenitor systems. As an illustration, we consider
here the constraints that can be derived from separating the host
galaxies into early and late types. On average, early-type galaxies
have their stars formed earlier than late-type galaxies, and this
difference, together with the time delay between progenitor formation
and short GRB outburst, leads to different burst rates in the two
types of hosts.  Presently available data suggest, but not yet prove,
that the local short GRB rate in early-type galaxies may be comparable
to that in late-type galaxies. This suggests that, unlike Type Ia
supernovae, at least half of the short GRB progenitors that can
outburst within a Hubble time have lifetimes greater than about 7 Gyr.
Models of the probability distribution of time delays, here
parametrized as $P(\tau)\propto \tau^n$, with $n \gtrsim -1$ are
favored. This apparent long time delay and the fact that early-type
galaxies in clusters make a substantial contribution to the local
stellar mass inventory can explain the observed preponderance of short
GRBs in galaxy clusters.
\end{abstract}


\section{Introduction} 
\label{sec:intro}

The progenitors of short duration, hard spectrum, gamma-ray bursts
(GRBs) are not yet well identified. Even with the recent localizations
of a handful of short-hard GRBs
\citep{Bloom06,Gehrels05,villa,Berger05,Fox05,Berger06,Berger07}, no
transient emission has been found that directly constrains the nature
of the progenitor system. The current view of most researchers is that
GRBs arise in a very small fraction of stars that undergo a
catastrophic energy release event toward the end of their evolution
\cite[e.g.,][]{moch93,ls76,elps89,nara92,rj99,kl98,rrr03,rrrd03,srj04,
Lee04,aloy,andrew,levan,LeeRR07}. Much of the
current effort is dedicated to understanding the different progenitor
scenarios and trying to determine how the progenitor and the burst
environment can affect the observable burst and afterglow
characteristics \citep[e.g.,][]{Lee05}. The lifetime of progenitors of
short bursts can be meaningfully constrained by properties of their
host galaxies (e.g., \citealt{Fox05,GalYam05}). We suggest here a
method to constrain the lifetime of the progenitors by making use of
the star formation histories of their host galaxies.

Based on the short burst afterglows localized so far \citep{Berger07},
two of the three relatively nearby ($z\lesssim 0.3$) events (GRB
050509b and GRB 050724) are plausibly associated with galaxies
exhibiting characteristic early-type spectra \citep{Berger05,Bloom06,
Gehrels05,Prochaska06}. In the other case (GRB 050709), the host
galaxy shows signature of a dominant stellar population with age of
$\sim$ 1Gyr \citep{Covino06} and it also exhibits strong emission
lines that indicate ongoing star formation
\citep{Covino06,Fox05,Prochaska06}. There is also independent support
that, at least two of the nine short bursts localized so far
\citep{Berger07} are associated with clusters of galaxies
\citep{Bloom06,Gladders05,Ped05}.  In contrast to what is found for
long-soft GRBs, for which all of the confirmed host galaxies are
actively forming stars \citep[e.g.,][]{trentham,Christensen04}, these
observations clearly signify that, like Type Ia supernovae, short GRBs
are triggered in galaxies of all types. What is more, it indicates
that there is a time delay between short burst occurrence and the main
epoch of star formation activity in the hosts, as determined by the
progenitor's lifetime.

In this paper, we present a method to deduce the lifetime of short GRB
progenitors from the burst rates in host galaxies of different types
that have distinct star formation histories. Limited by the accuracy
of how well we can separate their contribution to the star formation
history, we show, with this constraint in mind, how a large sample
host galaxies could be used to determine the lifetime distribution of
short GRB progenitors. Here we chose to separate the formation history
of stars residing in early- and late-type galaxies, but, as shown in \S
5, our study can be easily generalized to any set of subpopulations of
galaxies possessing distinct star formation histories. The layout is
as follows. In \S~2, we review the local stellar budgets in early- and
late-type galaxies. The total star formation history is subsequently
decomposed in \S~3 into a sum of the stars that assembled in today's
early- and late-type galaxies.  With the decomposition in place, in
\S~4, we show how the frequency of short GRBs in a well-defined sample
of host galaxies of different types can be used to constrain
the lifetime of the progenitors. Finally, in \S~5 we summarize our
results and outline future prospects. Throughout this paper, we assume
a spatially-flat $\Lambda$CDM cosmology with
$\Omega_m=1-\Omega_\Lambda=0.3$ and the Hubble constant $h=0.7$ in
units of $100{\rm km\, s^{-1}Mpc^{-1}}$.

\section{The Local Stellar Mass Inventory}
\label{sec:stellarmass}

The focus of this paper is the relatively local ($z\sim 0$) host
galaxy population, but our study can be easily generalized to any
redshift. The main idea is to use the difference in the star formation
history of galaxy sub-populations to probe the lifetime distribution
of short GRB progenitors. As an example, here we use the event rate of
burst residing in local early- and late-type galaxies.  On average,
early-type galaxies have their stars formed earlier than late-type
galaxies, and this difference, together with the delay time for short
GRB outburst, leads to different burst rates in these two types of
galaxies. As shown in \S~3, our method of inferring the star formation
history of a given population relies on tracing the assembly's history
of the stellar mass in galaxies at $z=0$.  For this reason, in this
section, we briefly review the local stellar mass budgets in galaxies
of different types.

With a universally applicable stellar initial mass function (IMF), the
stellar mass function (MF) of galaxies can be estimated from
well-defined samples of galaxies. Galaxy MFs have been measured based
on a few large galaxy redshift surveys
\citep[e.g.,][]{Kauffmann03,Bell03}.  For the calculations presented
here, we adopt the galaxy MF measured by \citet{Bell03} using a large
sample of galaxies from the Two Micron All Sky Survey (2MASS;
\citealt{Skrutskie06}) and the Sloan Digital Sky Survey (SDSS;
\citealt{York00}). The Bell et al.  catalog provides well-defined
samples of galaxies of different types as defined by either
the light concentration or the color of galaxies.

\begin{figure}
\plotone{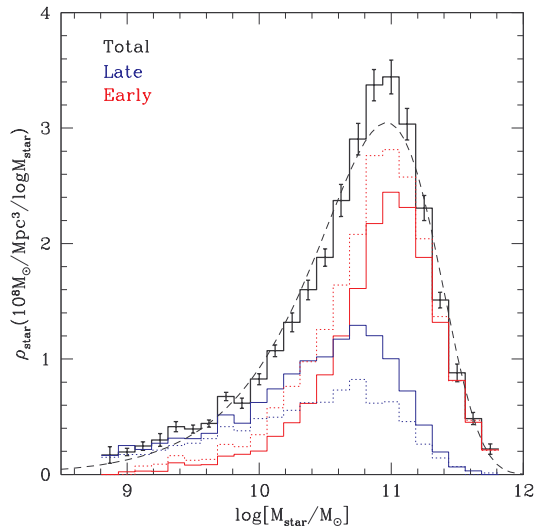}
\caption{
\label{fig:mf}
The local stellar mass density from galaxies of different
types. The histogram with errorbars is for galaxies of all types while
the dashed curve is a Schechter function fit to the stellar mass 
function of the entire galaxy sample. Red and blue histograms are for
early- and late-type galaxies, respectively. The early- and late-type
galaxies are divided according to their color ({\it dotted}) or light
concentration ({\it solid}).  This plot is based on the galaxy stellar
mass functions derived by \citet{Bell03}.  }
\end{figure}

In Figure~\ref{fig:mf}, we show the stellar MF, $\phi$, multiplied by
the stellar mass, $\Mstar$, for galaxies of different 
types. The quantity $\rhostar(\Mstar)=\Mstar\phi(\Mstar)$ is the
stellar mass density contributed by galaxies with stellar mass
$\Mstar$. The peak of $\rhostar$ for galaxies of all types comes from
$M^*$ galaxies, where $M^*\sim 10^{11} \Msun$ is the characteristic
stellar mass derived from the Schechter function fit.  The stellar
mass density distribution for late-type galaxies peaks at masses
slightly less than $M^*$ and has a tail extending to low stellar
masses.  Early-type galaxies have a narrow stellar mass density
distribution around $M^*$ and dominate the local stellar mass budget
above a few times $10^{10}\Msun$. Although early-type galaxies are far
less in total number than late-type galaxies, they are on average more
massive and thus make a larger contribution to the local stellar mass
density. The total stellar mass from early-type galaxies is about 1.3
(2.3) times that from late-type galaxies, if the classification of
galaxy types is based on light concentration (color).

If early- and late-type galaxies had similar star formation histories,
then the ratio of the short GRB rates in galaxies of different types
would simply be given by the ratio of the stellar mass density in
these galaxies.  However, the two types of galaxies have distinct star
formation histories -- on average, stellar populations in early-type
galaxies are older than those in late-type galaxies. For this reason,
the delay time between formation and the short GRB outburst plays an
important role in determining the burst rates in these two types of 
galaxies.

\section{The Star Formation History of Galaxies}
\label{sec:formationhistory}

The cosmic star formation history (SFH), which includes stars forming
in both early- and late-type galaxies, can be probed by different
techniques, including the rest-frame UV continuum emitted by
young stars, the infrared or sub-millimeter reprocessed radiation,
and line emission from star-forming regions (e.g.,
\citealt{Madau96,Giavalisco04, Perez05,Hopkins00}; see
\citealt{Hopkins04} and \citealt{Fardal06} for a summary and a
compilation of SFHs from different techniques).

Analytic fits to the SFH based on compiled star formation rate (SFR) 
at various redshifts
have been derived by many authors (e.g.,
\citealt{Hopkins04,Hopkins06,Cole01, Fardal06}). Here, as an
illustration, we use the analytic fit to the overall SFH given by
\citet{Fardal06},
\begin{equation}
\label{eqn:SFR_all}
{\rm SFR}_{\rm all}(z)=
\frac{p_1p_2p_3\rho_0(1+z)^{p_2}}{[1+p_1(1+z)^{p_2}]^{p_3+1}} H(z),
\end{equation}
where $\rho_0=9.0\times 10^8 \Msun {\rm Mpc}^{-3}$, $p_1=0.075$,
$p_2=3.7$, $p_3=0.84$, and $H(z)$ is the Hubble constant at redshift
$z$. The above fit assumes a diet Salpeter stellar initial mass
function\footnote{A diet Salpeter IMF is similar to the typical
  Salpeter IMF at the high mass end but differs at the low end by
  having less lower mass stars so that the total stellar
  mass content is 0.7 times that of the Salpeter IMF (\citealt{Bell01}).}  
(IMF), To
the first order, altering the IMF leads to an overall vertical shift
of the SFR. For example, adopting a Salpeter IMF, the SFR would be
about 40\% higher. What is more, even with a fixed IMF, the derived
SFR as a function of redshift shows considerable scatter. As a result,
the SFR in equation~(\ref{eqn:SFR_all}) may be scaled up and down by
about 0.2 dex \citep{Fardal06}.

In addition to the total SFH, we are also interested in knowing the
SFH for stars residing in galaxies of different types.  More
specifically, given the $z\sim 0$ early- and late-type galaxy
populations, our aim is to understand how the stars in these two
different types of galaxies were pieced together.  We, of course, only
need to know the SFH for galaxies of one type, as the SFH for the
other type can be easily derived by subtracting the known contribution
from the total SFH in equation~(\ref{eqn:SFR_all}).

Ideally, to determine the SFH of a given population of $z\sim 0$
galaxies, one would like to empirically derive their assembly history
by identifying their progenitor systems at different redshifts.
Alternatively, we can study the stellar population contents of $z\sim
0$ galaxies to infer their SFH. The former is not trivial, while
considerable efforts have been made to the latter through spectral 
synthesis modeling (e.g.,
\citealt{Heavens04,Panter06,CidFernandes05,CidFernandes06}), which
decomposes the observed spectra of galaxies into stellar populations
of a range of ages and metallicities. \citet{Heavens04} first
used this technique to infer the global SFH and that as a function of 
stellar mass. More recently,
\citet{Panter06} improved it to yield a better agreement with other
empirical determinations of the global SFH, which peaks at 
systematically higher redshifts than that in \citet{Heavens04}.  
Obviously, this method would improve with a better understanding of the 
systematic effects.

A more readily available method is to use galaxy formation models to
separate the SFH as a function of galaxy type. The accuracy of galaxy
formation models crucially depends on our understanding of the gas
physics and star formation as well as feedback processes. Current
galaxy formation models, although far from providing a perfect
description, can successfully explain many aspects of the observed
properties of galaxies. Hopefully, future improvements in galaxy
formation modeling would help provide a better description of the SFH
as a function of galaxy type.

With this in mind, we proceed separating the global SFH using a
particular galaxy formation model (with model uncertainties roughly 
accounted for).
It should be noted here that our method is applicable
to any two or more galaxy populations provided they have distinct
SFHs. This is clearly illustrated in \S~5, in which, for comparison,
an observationally motivated decomposition based on galaxy masses is
used instead.
 
\citet{DeLucia06} studied the formation history of elliptical galaxies
using a galaxy formation model based on the {\it Millennium
  Simulation} of the concordance $\Lambda$CDM cosmology
\citep{Springel05}. They calculated the average SFH of $z\sim 0$
elliptical galaxies of various stellar masses. If early-type galaxies
are identified here as ellipticals, we can then compute their average
star formation rate ($\Msun {\rm yr}^{-1} {\rm Mpc}^{-3}$) as a
function of redshift using
\begin{equation}
\label{eqn:SFR_early}
{\rm SFR}_{\rm early}(z)=\int d\Mstar \frac{dF(\Mstar,z)}{d\Mstar}
                   \Mstar \phi_E(\Mstar),
\end{equation} 
where $\phi_E(\Mstar)$ is the $z=0$ stellar MF of early-type galaxies
derived by \citet{Bell03}, and $dF(\Mstar,z)/d\Mstar$ (in units of
$\Msun {\rm yr}^{-1} \Mstar^{-1}$) is the average SFR per stellar mass
for a $z=0$ elliptical galaxy of stellar mass $\Mstar$ (obtained from spline 
interpolations between the curves presented in \citealt{DeLucia06}). 

The application of equation~(\ref{eqn:SFR_early}) combines the
empirically inferred galaxy stellar mass function and the
theoretically predicted SFR. It requires the galaxy samples derived by
both observation and theory to be consistent.  Although in principle
this is not a problem, since any sophisticated galaxy formation model
would give detail predictions for a well defined galaxy population,
the available data and models used here are based on slightly
different definition rules, which introduces uncertainties in the
determination of the SFH of early-type galaxies. In \citet{Bell03},
galaxies are classified through either color or light concentration,
which is shown to be only a crude discriminant between early- and
late-type galaxies \citep{Strateva01}, and, as a result, the
early-type galaxies in \citet{Bell03} include galaxies earlier than
Sa. \citet{DeLucia06} split galaxies in their formation model
according to the predicted $B$-band bulge-to-disk light ratio and
classify them as ellipticals provided the bulge light is at least 70\%
of the total light. For this reason, the SFH of the observed local
early-type galaxies as derived from the stellar-mass dependent SFH of
the modeled ellipticals would tend to shift toward slightly earlier
formation times. In addition, in the galaxy formation model used here,
stars are produced from cooled gas and a stellar IMF is adopted only
for the purpose of providing spectroscopic predictions. Because of
this, a comparison between this IMF and the one used to derive the
observed galaxy stellar mass function is not meaningful. Instead, the
key questions are whether the IMF assumed to derive the observed
galaxy mass function is close to the ``true'' one and whether the star
formation efficiency in the galaxy formation model is
accurate. Inaccuracy in either one would lead to a relative amplitude
change in the SFHs.

\begin{figure*}
\plotone{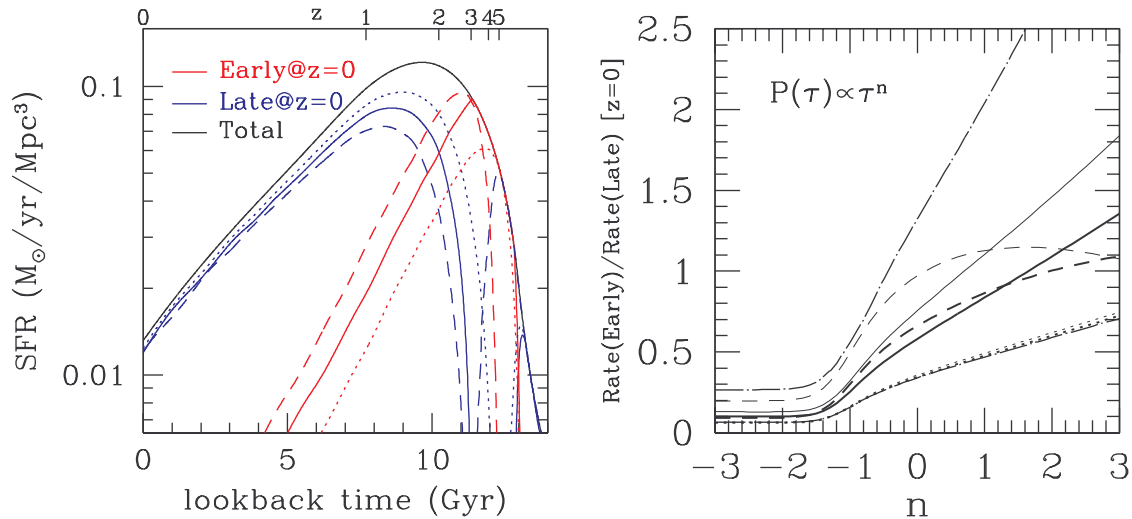}
\caption[]{
\label{fig:SFR}
{\it Left panel:} Star formation histories for $z\sim 0$ galaxies of
different types. The total star formation history ({\it
black}) is decomposed into two parts --- formation histories of stars
that are now in the $z\sim 0$ early- ({\it red}) and late-type
galaxies ({\it blue}). Solid, dotted and dashed red and blue curves
result from different treatments of the SFH of early-type galaxies
(see text), which give a clear representation of the uncertainties
associated with the assumed SFH decomposition. The curves shown here
are for early- and late-type galaxies as defined by the concentration
of their light profile (corresponding to the solid histograms in
Fig.\ref{fig:mf}).  {\it Right panel:} The ratio of short GRB rates in
early- and late-type galaxies at $z\sim 0$ as a function of the index
$n$ of the progenitor lifetime distribution $P(\tau)\propto
\tau^n$. The thick solid, dotted and dashed curves correspond to the
different SFH splits shown in the left panel. The thin curves, on the
other hand, are for the same SFH decomposition but for galaxy types
classified based on color instead of light concentration. The two
dot-dashed curves give a fair representation of the uncertainties in
the determination of the burst ratio, which arise primarily from the
different methods and adjustments used here to decompose the SFH (see text).}
\end{figure*}

The left panel of Figure~\ref{fig:SFR} shows an example of the SFH
decomposition. Although the primary aim of the SFH decomposition
selected in this study is to illustrate the applicability of our
method, we pay particular attention to the uncertainties associated
with it.  Several variants of the SFH decomposition are constructed to
test the robustness of the results. Each version has a different
way to adjust the derived SFR in early-type galaxies with respect to
the assumed overall SFH. The amplitude and/or position of the derived
rate is adjusted using one of the following methods: (i) the derived
SFR is selected to be equal to the overall SFH whenever the predicted
rate lies above it (red solid curve in Fig.~\ref{fig:SFR}), (ii) the
overall amplitude of the derived SFR is systematically lowered so that
it lies just below the overall SFH (red dotted curve in
Fig.~\ref{fig:SFR}), and (iii) the derived SFR is arbitrarily shifted
to low redshifts so that it both lies below the overall SFH and peaks
at a redshift that is not lower than the overall SFH (red dashed curve
in Fig.~\ref{fig:SFR}). In addition, the overall SFH is allowed to
change by 0.2 dex in order to account for the uncertainties associated
with its determination \citep{Fardal06}.  The above variants are
applied every-time the overall SFH is altered.

Finally, we adopt the two definitions of early-type galaxies in
\citet{Bell03} based on color and concentration, respectively. As a
whole, the above procedure takes into account both the inaccuracy of
the galaxy formation model predictions and the uncertainties in the
global SFH as inferred from observations, as well as the ambiguity of
the earl-type definition criteria. These uncertainties are propagated
self consistently when estimating the lifetime distribution of short
GRB progenitors (\S~\ref{sec:lifetime}).  The region enclosed by the
dashed-dotted curves in the right panel of Fig.~\ref{fig:SFR} shows
the combined effects on the derived lifetime distributions of short
GRB progenitors when adopting the three variants discussed above,
allowing a 0.2 dex change in the overall SFH, and considering the two
definitions of early-type galaxies.

As a representative example, in the left panel of
Figure~\ref{fig:SFR}, we plot the SFH in equation~(\ref{eqn:SFR_all})
and the SFH of the $z=0$ early-type galaxies defined through color, as
black and red curves, respectively. The different line types for the
SFH of early-type galaxies are for the various adjustments listed
above. In general, the derived average SFHs of the $z=0$ early-type
galaxies peak at a look-back time of about $\sim$10--12 Gyr and
subsequently decline toward low redshifts. The shape of this
early-type SFH is quite consistent with that of the fossil bulge model
fit in \citet{Nagamine06} and that of SFH of galaxies with stellar
mass above $10^{11}\Msun$ inferred in \citet{Panter06}.

The average SFH for the $z=0$ late-type galaxies, computed by
subtracting (\ref{eqn:SFR_early}) from (\ref{eqn:SFR_all}), is shown
as blue curves in the left panel of Figure~\ref{fig:SFR}.  The average
SFR for the $z=0$ late-type galaxies has a steep rising starting at
lookback times of $\sim$11 Gyr, subsequently peaking at $\sim$8--9
Gyr, and slowly declining thereafter.  The SFR becomes increasingly
dominated by late-type galaxies from lookback times of $\sim$9 Gyr to
the present epoch.

We note here that the overall SFH in equation~(\ref{eqn:SFR_all}) has
a slightly higher amplitude than the predicted early-type SFR at
lookback times greater than 12 Gyr, which gives rise to a modest second
peak in the late-type SFR at high redshifts. This may be interpreted
as a contribution from local late-type galaxies to the SFR at these
early epochs, which could be related to the formation of the bulges of
these galaxies. It may also be an artifact associated with the
inaccuracy of the model, which we consider to be more likely.  We find
that even if taken at face value, this early contribution to the
formation of stars in local late-type galaxies leads only to a less
than 10\% increase in the inferred short GRB burst ratio in the two
types of galaxies (\S~\ref{sec:lifetime}). For this reason, we leave
the SFR for late-type galaxies at these epochs as those given by the
decomposition and do not perform any correction.

\section{Constraints on the Lifetime of Short GRB Progenitors}
\label{sec:lifetime}

The short GRB rate is a convolution of the SFR and the distribution of
time delays between formation and outburst \citep[e.g.,][]{Piran92}
\begin{equation}
\label{eqn:rate}
R_i(z) = C\int_0^{t(z)} d\tau\, {\rm SFR}_i(t-\tau)P(\tau),
\end{equation}
where the subscript $i$ denotes the type of galaxies in consideration,
$P(\tau)$ is the probability distribution of the time delay $\tau$,
and $C$ is a normalization constant.  In principle, details of the
star formation process may be related to the assembly history of
galaxies. Therefore, the distribution $P(\tau)$ and the normalization
constant $C$ could be different for different types of galaxies. We
make the simple assumption here that this dependence is weak so that
the same distribution and normalization is used for all types of
galaxies.

The distribution $P(\tau)$ of the time delay for short GRB is not yet
well understood. It can be constrained using, for example, the burst
rate as a function of redshift
\citep[e.g.,][]{nakar06,guetta05,Ando04}. The luminosity function of
short GRBs and its redshift evolution used in this method are not well
determined at present. To avoid such complications, here we propose to
use the local rates of short GRBs in different types of galaxies as an
alternative and complementary method to constrain the probability
distribution $P(\tau)$. As an illustration, we adopt a simple
parameterization, $P(\tau)\propto \tau^n$, and calculate the ratio of
rates of short GRBs at $z\sim 0$ in early- and late-type galaxies as a
function of $n$.  The right panel of Figure~\ref{fig:SFR} shows the
results of such exercise. The thick solid, dotted, and dashed curves
are for three different treatments of the early-type SFR (see
\S~\ref{sec:formationhistory}), corresponding to the three different
curves in the left panel. Here galaxy types are defined by their light
concentration. The three thin curves in the right panel have been
calculated using a similar treatment but are for early-type galaxies
as defined by their color. The two dot-dashed curves give the maximum
range of uncertainties in the resultant burst rate ratio, as they give a fair
representation of all the uncertainties involved in separating the SFH
as discussed in \S~3.

For larger $n$, the distribution $P(\tau)$ is weighted more toward
longer time delays, so that early-type $z\sim 0$ galaxies, which on
average form their stars earlier than late-type $z\sim 0$ galaxies,
are more likely to host short GRBs. For smaller (negative) $n$, short
GRB progenitor systems with shorter time delays would dominate and one
would be more likely to find them in late-type galaxies. For
$n\lesssim-3/2$, short GRB progenitors would be those with very short
time delays so that the short GRB rate ratio is basically determined
by the ratio of the $z\sim 0$ SFRs in early- and late-type
galaxies. However, the amplitude of the plateau at $n\lesssim-3/2$ in
the right panel of Figure~\ref{fig:SFR} should be taken cautiously,
because it depends on how one interpolates the SFR in early-type
galaxies derived by \citet{DeLucia06} to $z\sim 0$. For this reason,
the method proposed here becomes less useful for time delays that are
short when comparing to the mean age of stars residing in early-type
galaxies. Clearly, as we argue in what follows, current observational
constraints for short GRBs are such that uncertainties in the
determination of the amplitude of the plateau at $n\lesssim-3/2$ are
irrelevant.

Our calculation is based on $z\sim 0$ galaxy populations. Of the nine
well localized short bursts, only three hosts, two elliptical galaxies
and one star-forming galaxy, reside at $z\lesssim 0.3$. 
Therefore, the face value of the burst ratio in {\it local} early-
and late-type galaxies is $\sim 2$ provided there are no
selection effects that make short GRBs more likely to be detected in a
galaxy of a given type (e.g., because of the low density of the
interstellar medium one may expect the GRB afterglow in elliptical
galaxies to be faint, e.g., \citealt{Belczynski06}). The error-bar
on this observed ratio is not readily assigned because of the small number
statistics. However, even with a conservative estimate of the ratio 
as $\sim 2\pm 1.5$, the analysis presented here alludes to a
probability distribution with $n\gtrsim-1$, thus favoring long delay
times.  This is in contrast to what is found in Type Ia supernovae,
for which the relatively low frequency in early-type galaxies
\citep[e.g.,][]{sidney,Mannucci05} yields $n\lesssim -1$.  

\begin{figure}
\plotone{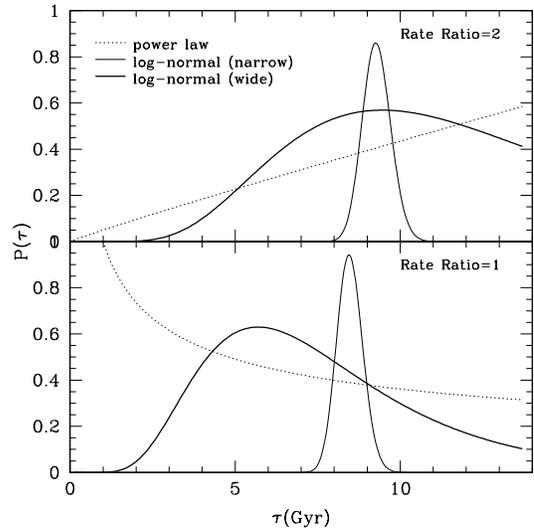}
\caption[]{
\label{fig:ptau}
Illustration of different functional forms of $P(\tau)$ that give rise
to the same observed short burst rate ratio in early- and late-type galaxies.
If $P(\tau)$ is assumed to have a log-normal form, there are a series of
combinations of the mean and the width that can explain a given
observed burst rate ratio, here assumed to be either 1 ({\it bottom panel})
or 2 ({\it top panel}). A narrow and a wide log-normal distributions
are shown.
}
\end{figure}

If the distribution $P(\tau)$ is assumed to have a log-normal form, there 
are a series of combinations of the mean and the width that can explain 
the observed burst rate ratio in early- and late-type galaxies at $z\sim 0$. 
We illustrate the underlying powerlaw and two cases of log-normal $P(\tau)$
distributions for burst ratio of one and two, respectively, in 
Figure~\ref{fig:ptau}, using the SFH split corresponding to the top curve 
in the right panel of Figure~\ref{fig:SFR}. For a narrow log-normal 
distribution, it just peaks at $\tau_0$ when the ratio of the SFRs in two 
types of galaxies at a lookback time $\tau_0$ equals the burst ratio. A 
wide log-normal
distribution and the corresponding powerlaw distribution have a general 
similarity with each other, but they diverge from each other at small
time delays if the powerlaw index is negative, which indicates
that the method has a poor constraint in $P(\tau)$ at the low 
$\tau$ end. For a burst ratio of unity, we find $n>-0.45$. If we take 
13.7 Gyr as an upper limit for the delay time, which allows short GRB 
progenitors to outburst within a Hubble time, a distribution $P(\tau)$ 
with $n=-0.45$ implies that about half of progenitors of short GRBs should 
have lifetimes longer than 3.9Gyr. If at lower $\tau$, $P(\tau)$ does not 
rise as steeply as a powerlaw, this number becomes higher --- for example,
with a cutoff in $P(\tau)$ at $\tau<$2Gyr, the number becomes 6.7Gyr.
For a wide (narrow) log-normal distribution, a burst ratio of order 
unity implies that about half of progenitors of short GRBs should 
have lifetimes longer than 7.0 (8.5) Gyr. 

Obviously, the above calculation is only sketchy and should be taken
with a grain of salt
at present, although, given the
uncertainties assigned to the modeling, the lower bound of $n$ seems
rather robust. Constraints should improve as more host galaxies of
short GRBs are detected and the SFH separation method is improved. In
order to reach a 10\% accuracy in the the burst ratio, observations of
some 100 local short GRB host galaxies are needed. If the split of the
SFHs could be well determined, an associated uncertainty $\Delta n\sim
0.3$ in the time delay distribution could be easily achieved. If the
realistic SFH separation lead to a 20\% uncertainty in the
predicted burst rate ratio, the constraints on $n$ would be
degraded to $\Delta n\sim 0.9$. On the other hand, as suggested by our
analysis, it is possible to set a very reliable lower bound on $n$.
In brief, for a more precise constraint to be derived on the time
delay of short GRB progenitors using the method proposed here, one
needs a larger sample of local short GRB host galaxies to be observed,
the split of the SFH to be better determined, and the classification
scheme of GRB host galaxies to be consistent with the assumed SFH
split.

\section{Discussion}

\begin{figure*}
\plotone{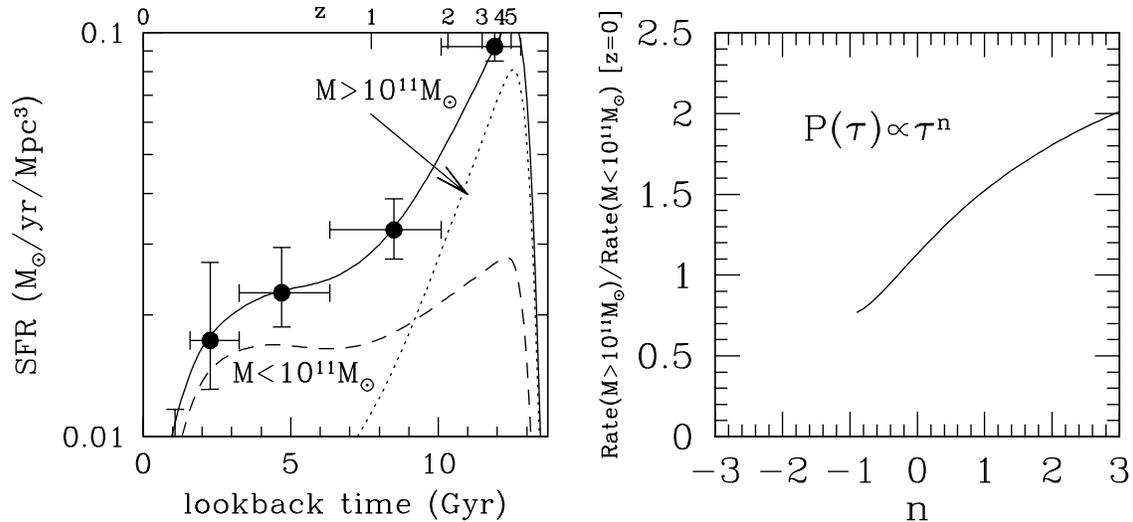}
\caption[]{
\label{fig:ratio}
Similar to Fig.~\ref{fig:SFR}, but the SFH is decomposed into those
for galaxies with different stellar masses. This decomposition is
based on the results in \citet{Panter06}, where the SFHs of local,
individual galaxies are inferred from modeling their spectra and are
then subsequently grouped according to the present stellar mass
content. In the left panel, the solid curve shows a spline fit to the
overall SFH inferred by \citet{Panter06}, and the dashed and dotted
curves give the spline fits to the SFRs of galaxies with stellar mass
below and above $10^{11}\Msun$, respectively. For lookback time larger
than $\sim$12.5 Gyr, a cutoff profile is introduced. In the right
panel, the ratio of short GRB rates in galaxies with stellar mass
above and below $10^{11}\Msun$ is plotted as a function of the index
$n$ of the progenitor lifetime distribution $P(\tau)\propto \tau^n$.
The purpose of the figure is to illustrate that the method presented
in this paper can be generalized to any sub-populations of galaxies
that have distinct SFHs (see text).  }
\end{figure*}

In the local universe, based on the stellar mass functions of galaxies,
about 55--70\% of the stellar mass is in early-type galaxies, and,
from the SFH we inferred, the corresponding stars mainly formed about
9 Gyr ago.  Two of the three host galaxies of local short GRBs
($z\lesssim 0.3$) are associated with old and massive galaxies with
little current or recent star formation, which makes it unlikely that
short bursts are associated with massive stars. Presently available
data suggests, but not yet prove, a long time delay between the
formation of the progenitor system and the short GRB outburst --- for
progenitors that can outburst within a Hubble time, about half of them
have lifetime longer than $\sim$7 Gyr. It is fair to conclude that,
based on the current host galaxy sample, the progenitors of short GRBs
appears to be longer lived than those of Type Ia
supernovae. \citet{Fox05} also reach similar conclusion by arguing
that Type Ia supernovae occur more frequently in late-type,
star-forming galaxies. Simply based on a comparison of Hubble types
between short GRBs and Type Ia supernovae host galaxies,
\citet{GalYam05} argue that the delay time of short GRBs should be
several Gyr even if the Type Ia supernovae delay time is as short as
$\sim 1$ Gyr.

The lifetime of the progenitor systems is estimated here by using the
SFH of elliptical galaxies from a galaxy formation model. This allow
us to separate the early- and late-type galaxy contributions to the
overall cosmic SFH.  It would, however, be more self-consistent to
infer SFHs of different types of galaxies by modeling the observed
spectra with stellar population synthesis models. In either case, the
uncertainty in the derived SFHs should be folded into the errors
derived by this method for the distribution of time delays of short
GRB progenitors.

In our calculation, different definitions of early-type galaxies (by
color or by light concentration) introduce uncertainties in the
resultant SFH. This, in principle, would not be a problem since we can
choose to use the same definition for identifying the short GRB host
galaxy type. The method proposed here is not limited to a separation
between early and late galaxies. As long as galaxies are divided into
two (or more) sub-populations that have distinct star formation
histories, they can be used to constrain the lifetime distribution of
GRB (or Type Ia supernova) progenitors, following the same reasoning presented
here.  For example, \citet{Shin06} recently applied our approach to
field and cluster elliptical galaxies. Another example would be to
divide galaxies according to their stellar mass. \citet{Panter06}
present SFH of local galaxies as a function of galaxy stellar mass by
modeling their spectra. Low mass galaxies on average form their stars
later than high mass galaxies, which can be used to constrain
$P(\tau)$ if the stellar mass of GRB hosts can be obtained. As an
illustration, in Figure~\ref{fig:ratio}, we show the separation of the
SFR of galaxies with low and high stellar masses (below or above
$10^{11}\Msun$) using results by \citet{Panter06}. The shape of the
overall SFH used in \citet{Panter06} is slightly different from that
in Figure~\ref{fig:SFR}, which may reflect some systematics in the
derivation of the SFH based on the fossil record.  If the systematics
can be well controlled, this approach can become even more powerful than
the one used here by inferring SFHs of individual GRB host galaxies.
A maximum likelihood method can then be used to constrain
$P(\tau)$ based solely on the SFH of host galaxies, which puts our
proposed method to an extreme --- dividing the galaxy populations into 
individual galaxies. Furthermore, instead of assuming a functional 
form of $P(\tau)$, such an application to a large number of individual 
host galaxies may allow constraints on a non-parametric form of the 
distribution [i.e., constraining $P(\tau)$ in different $\tau$ bins]. 

In this paper, we
limit our study to $z\sim 0$ galaxies, but the method can be easily
generalized to galaxy populations at any redshift provided that one
can accurately infer their SFHs (e.g., through spectral synthesis).
The observational task can be minimized by focusing on the observations
of host galaxies of individual short GRBs.  However, for applications
at high redshift to be useful, the luminosity function of short GRBs
needs to be understood.

Throughout this paper we have assumed the same lifetime distribution
of short GRB progenitors in both early- and late-type
galaxies. However, star formation processes in these two types of
galaxies may not be identical. For example, elliptical galaxies can
form by the merging of two gas-rich galaxies
\citep[e.g.,][]{mihos94}. Many globular clusters can form in the
merging process \citep{schweizer}, which could enhance, for example,
the fraction of binary progenitors and also change the lifetime
distribution (e.g., \citealt{Grindlay06}).  The magnitude of this kind
of effect on the GRB progenitors is a formidable challenge to
theorists and to computational techniques. It is, also, a formidable
challenge for observers, in their quest for detecting minute details
in extremely faint and distant sources.

At least two short GRB host galaxies are found in cluster
environments. There may exist a selection bias of detecting short GRBs
in a dense medium \citep{Bloom06}. To study the association of short
GRBs with clusters, it would be useful to separate the stellar mass
function into that for field galaxies and that for cluster galaxies in
addition to early- and late-type galaxies. More promising for the
immediate future, the preponderance of cluster environments can be
investigated observationally. Important information may be gained by
studying the local stellar mass inventory shown in
Figure~\ref{fig:mf}.  Approximately 50\% of the stellar mass contents
in early-type galaxies are in galaxies with $M_{\rm
star}>10^{11}\Msun$ that typically reside in clusters.  Since it is
likely that in clusters galaxies shut off their star formation process
early on, a long progenitor lifetime further increases the tendency
for short GRBs to happen in cluster galaxies. It is fair to conclude
that the observed preponderance of cluster environments for short GRBs
is consistent with an old stellar population that preferentially
resides in early-type galaxies.

Detailed observations of the astrophysics of individual GRB host
galaxies may be essential before stringent constraints on the lifetime
of short GRB progenitors can be placed. If confirmed with further host
observations, this tendency of short GRB progenitors to be relatively
old can help differentiate between various ways of forming a short
GRB.

\acknowledgments 
We thank J. Bloom, N. Dalal, W. Lee, D. Pooley, and J. Prochaska for 
helpful conversations and John Beacom for useful comments.  We thank the
referee for detailed comments that improved this paper. We acknowledge 
the support of NASA through a Hubble (ZZ) and Chandra (ER-R) Postdoctoral 
Fellowship awards HF-01181.01-A and PF3-40028, respectively.

\end{document}